\documentclass[twocolumn,prl,showpacs,preprintnumbers,amsmath,amssymb,dvips,epsfig]{revtex4}
\usepackage{graphicx}
\usepackage{dcolumn}
\usepackage{bm}
\usepackage{textcomp}

\begin{document}



\title{Dynamical Self-assembly during Colloidal Droplet Evaporation Studied by {\em in situ} Small Angle X-ray Scattering }


\author{Suresh Narayanan, Jin Wang}
\affiliation{Advanced Photon Source, Argonne  National Laboratory,
Argonne, IL 60439}
\author{Xiao-Min Lin}
\affiliation{Materials Science Division, Chemistry Division and
Center for Nanoscale Materials, Argonne National Laboratory,
Argonne, IL 60439}


\date{\today}

\begin{abstract}
The nucleation and growth kinetics of highly ordered nanocrystal
superlattices during the evaporation of nanocrystal colloidal
droplets was elucidated by {\em in situ} time resolved small-angle
x-ray scattering. We demonstrated for the first time that
evaporation kinetics can affect the dimensionality of the
superlattices. The formation of two-dimensional nanocrystal
superlattices at the liquid-air interface of the droplet has an
exponential growth kinetics that originates from interface
"crushing".
\end{abstract}

\pacs{78.67.Bf, 61.46.+w, 81.16.Dn}

\maketitle



The microscopic mechanism for self-organization of
ligand-stablized nanocrystals has been well established
\cite{Oha,Wan}. The interparticle van der Waals interaction
provides the attractive force to induce the self-assembly, whereas
the steric interaction due to the ligand interdigitation provides
the balancing force to create stable structures. However, the
macroscopic patterns created by such interactions can vary
dramatically under different experimental conditions. Even for a
simple case of evaporating a colloidal droplet of nanocrystals on
a surface, a variety of patterns have been reported, ranging from
ordered two-dimensional(2D), three-dimensional(3D) nanocrystal
superlattices (NCSs) \cite{Mur2,Mot,Lin1}, fractal-like aggregates
\cite{Ge,Tan} to more percolated networks \cite{Sto,Mai,Mor}. A
common notion is that the self-assembling process occurs at the
liquid-substrate interface \cite{Mur3}, where the competing effect
from the diffusion of nanocrystals along the substrate and solvent
dewetting could lead to pattern formation far away from
equilibrium \cite{Rab}.  Having such vast different macroscopic
structures of nanocrystal assembly hinders the exploration of
their physical properties \cite{Mur2,Rem, Par}, as well as
creating obstacles for device applications \cite{Mur3,Wel}.  It is
therefore essential to further understand and control the
nanocrystal self-assembly mechanism.

Recently, it has been shown that highly ordered 2D NCSs with
domain size up to tens of microns, can be formed on a silicon
nitride substrate by evaporating a nanocrystal colloidal droplet
with slight excess amount of dodecanethiol ligand molecules in the
solution \cite{Lin1}.  Figure 1a shows a transmission electron
microscopy (TEM) image of a region of such highly ordered NCSs.
The high degree of ordering and large domain size indicates the
formation of NCSs most likely occurs through nucleation and growth
mechanism. It is difficult, however, to explain the high
crystallinity of the superlattices through a mechanism in which
self-assembly occurs on the substrate at the last instance when
the solvent dewets the surface. In this letter, we report an {\em
in situ}, non-intrusive, small angle x-ray scattering (SAXS)
measurement to elucidate the formation of NCSs as the colloidal
droplet evaporates. We show that, contrary to the previous notion,
these highly ordered 2D superlattices are formed at the liquid-air
interface of the liquid droplet during the solvent evaporation.
Changing the solvent evaporation rate can lead to either 2D or 3D
NCS formation using the same colloidal nanocrystals.

We used monodispersed gold nanocrystals with an average diameter
of 7.5 nm and 5.8 nm in two different experimental runs
\cite{Lin3, Stoe}, respectively, and with the particle number
concentration adjusted to be approximately $10^{13}
\textrm{mL}^{-1}$, just enough to form a monolayer. The volume
concentration of thiol was 0.63\%. The SAXS experiments were
conducted at the 1-BM beamline of the Advanced Photon Source
(APS). 10 $\mu$l of colloidal solution was deposited onto a
rectangular silicon nitride substrate ($3\times 4
\textrm{mm}^{2}$). The incident synchrotron x-ray beam was
monochromatized to 8.0 keV by a double-crystal monochromator while
two sets of horizontal (H) and vertical (V) slits were used to
define the beam size to 0.2 (H) $\times$0.2 (V)$\textrm{mm}^{2}$.
We chose a laboratory coordinate system as shown in Figure 1b so
that the incident x-ray beam is along the z-direction and the
substrate surface is in the y-z plane. The samples were initially
aligned using the substrate as a reference, thus the top surface
of the substrate was centered in the x-ray beam along the
x-direction. During the experiment, the substrate was slightly
titled with an angle of $0.3$\textdegree\hspace{1pt} relative to
the incident beam. The scattered x-rays passed through a helium
flight path
 and were collected by either an image plate (IP) or a charge-coupled device (CCD) detector.
\begin{figure}
\scalebox{0.35}{\includegraphics{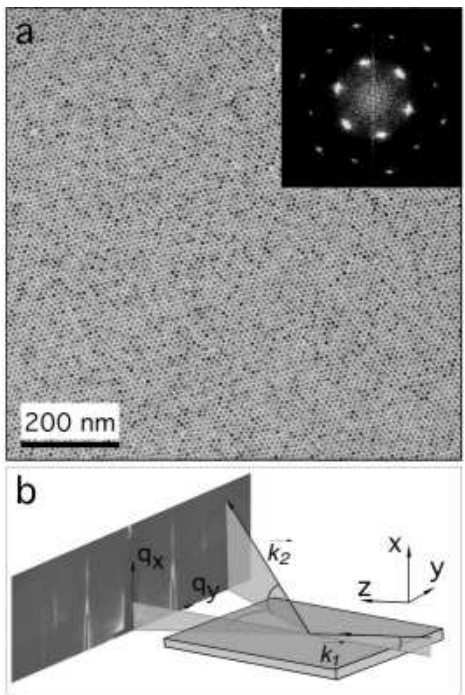}}
\caption{\label{fig1}Characterization of 2D nanocrystal
superlattices formed by colloidal droplet evaporation: (a)
Transmission electron microscopy (TEM) image of a highly ordered
gold nanocrystal monolayer. Inset shows Fourier transformation of
the superlattice. (b) Schematic diagram of scattering geometry
used for the {\em in situ} SAXS measurement. The dimension of the
substrate is exaggerated in order to clearly illustrate the
experimental setup.}
\end{figure}

The dynamical self-assembling process can be observed by
positioning the incident beam at the bottom center of the droplet
to accommodate the changing meniscus of the droplet as it
evaporates. Figure 2 shows a typical time evolution of the SAXS
patterns after the droplet is deposited on the substrate. The
droplet is evaporated in air at room temperature (23\textcelsius)
with the mass of the droplet decreasing approximately at 0.66mg
per minute. The initial thickness of the droplet is typically more
than 2 mm. No visible scattering pattern is observed in the first
two minutes indicating that, initially, there are no ordered
superstructures in the droplet (Figure 2a). After that, an
elliptically shaped diffraction ring is observed (Figure 2b). The
section of the scattering ring near $q_{y}=0$ gradually becomes
more diffuse in the ${q_{x}}$-direction and eventually disappears,
while the intensity of the scattering corresponding to the
in-plane long-range order increases dramatically (Figures 2c-2d).
The scattering pattern in Figure 2d remains unchanged for hours
during which a thin liquid film, less than 100 $\mu$m in
thickness, with a high concentration of dodecanethiol remains on
the substrate surface.   To illustrate the kinetics of the
nanocrystal array growth, Figure 2e shows the experimental time
evolution of the intensity of the (10) powder diffraction peak at
$q_{x}=0$, plotted in a semi-natural log format. The scattering
intensity increases exponentially with time before becoming
saturated after 8 minutes (data not shown). When the liquid film
is completely dried and the NCSs are deposited directly on the
substrate, the downward scattering is blocked by the substrate in
the grazing incidence geometry \cite{Dos,Sup1}. In contrast, the
scattering patterns in the liquid film (Figure 2d) is highly
symmetric in the vertical direction ($q_{x}$), indicating the 2D
NCSs are formed at the liquid-air interface which is elevated
above the substrate.  The self-assembly of nanocrystals at the
liquid-air interface was further confirmed by using a wider x-ray
beam measured to be 1.5 (H)$\times$ 0.2 (V)$\textrm{mm}^{2}$,
purposely positioned near the edge of the droplet \cite{Sup2}.
Bragg diffraction patterns from the 2D NCSs were observed rotating
along with the direct reflection of meniscus surface as the
droplet flattens due to evaporation. We should emphasize that,
unlike experiments carried out on the Langmuir trough \cite{Sea},
no aqueous subphase was used in our experiments. Both the
nanocrystal ligand and the toluene solvent are hydrophobic in
nature.

\begin{figure}
\scalebox{0.35}{\includegraphics{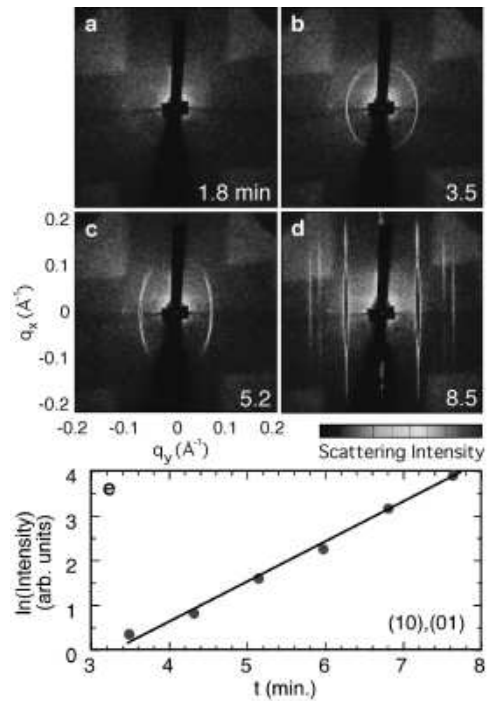}}
\caption{\label{fig2}(a-d){\em In situ} SAXS patterns of 2D NCSs
formation during the droplet evaporation.  Time in unit of minute
is in reference to the deposition of colloid droplet. The average
diameter of the nanocrystals is 7.5 nm determined by TEM. Cross
bar feature in each frame is the shadow image of the beam stop.
(e) Time-evolution of the normalized scattering intensity of the
monolayer powder diffraction peak for (10) powder diffraction peak
(solid line is a linear fit to the data).}
\end{figure}

A comprehensive understanding of the formation of the 2D NCS
monolayer, including the early stage scattering ring, can be
achieved by comparing the simulated SAXS patterns during the
entire evaporation process with the experimental data. The initial
elliptically-shaped x-ray scattering patterns as well as its
evolution can be modelled by the "powder" diffraction of 2D
superlattice domains with varying x-ray incident angles. As shown
in the schematic cross section diagram in Figure 3A, the x-ray
incident angle $\alpha$ at the liquid-air interface varies during
the droplet evaporation (while the edges of the droplet are pinned
at the substrate). To model the x-ray scattering patterns from the
2D lattices with their lattice plane orientation changing with
time, we define a sample coordinate system $(x',y')$
 that is attached to the lattice plane as shown in Figure 3a.  In a kinematical approach of x-ray diffraction \cite{War}, the averaged powder diffraction intensity from many 2D lattice domains in
 the reciprocal space is distributed in a circular ring defined by $\sqrt{q_{x'}^{2}+q_{y'}^{2}}=|\bf{q_{h,k}}|$, where ($q_{x'},q_{y'}$) are the momentum transfer (MT) components
 in the sample coordinates which satisfies the Bragg diffraction
condition defined by the scattering vector $\bf{q_{h,k}}$.  The
scattering patterns recorded by the 2D detectors in the laboratory
space are related to the circular powder
  ring in the sample space by a linear transformation between the two coordinates.  Therefore the Bragg scattering condition in the laboratory coordinate is
  described approximately by $\sqrt{(q_{x}\sin\alpha)^{2}+q_{y}^{2}}=|\bf{q_{h,k}}|$, where ($q_{x},q_{y}$) are the MT components in laboratory coordinate.
  The short axis of the ellipse is in the $q_{y}$-direction
  and has a magnitude of $|\bf{q_{h,k}}|$,
  whereas the long axis is in the $q_{x}$-direction and has a magnitude of $|\bf{q_{h,k}}|/\sin\alpha$ . For example, the elliptical scattering pattern in Figure 2b indicates,
  at 3.5 minutes into the evaporation process, the 2D NCSs formed at the interface which is orientated about 52.4\textdegree\hspace{1pt} with respect to the incoming x-ray beam.
  The complete scattering intensity is the product of the lattice structure factor and the form factor of a spherical nanocrystal expressed as
  $|F(q)|=|[\frac{3(\sin(qr)-qr\cos(qr))}{(qr)^3}]|^{2}$,
  where $q=\sqrt{q_{x}^{2}+q_{y}^{2}+q_{z}^{2}}$ and $r$ is the radius of the sphere.  As the liquid droplet evaporates, the scattering geometry approaches a grazing
  incidence angle with respect to the sample ($\alpha=0.3$\textdegree).
   In this case, the scattering intensity distribution along the $q_{y}$-direction corresponds to the structure factor of the in-plane lattice and that along $q_{x}$ reveals the form factor
   of the individual nanocrystal. Figures 3b-e show the simulation results at different stages of evaporation, based on a 2D hexagonal close-packed monolayer of nanocrystals
with an average diameter of 7.5 nm (polydispersity of 7\%) and a
10.3 nm center-to-center spacing.  Because the incident angle can
be deduced from the simulation, it is possible to obtain the NCS
scattering intensity normalized by the x-ray beam footprint size.
This allows a precise determination of the self-assembly kinetics.
The x-ray scattering intensity data shown in Figure 2e is found to
be proportional to the integrated intensity after being normalized
by the changing x-ray footprint size on the sample due to the
variation of x-ray incident angle.

\begin{figure}
\scalebox{0.40}{\includegraphics{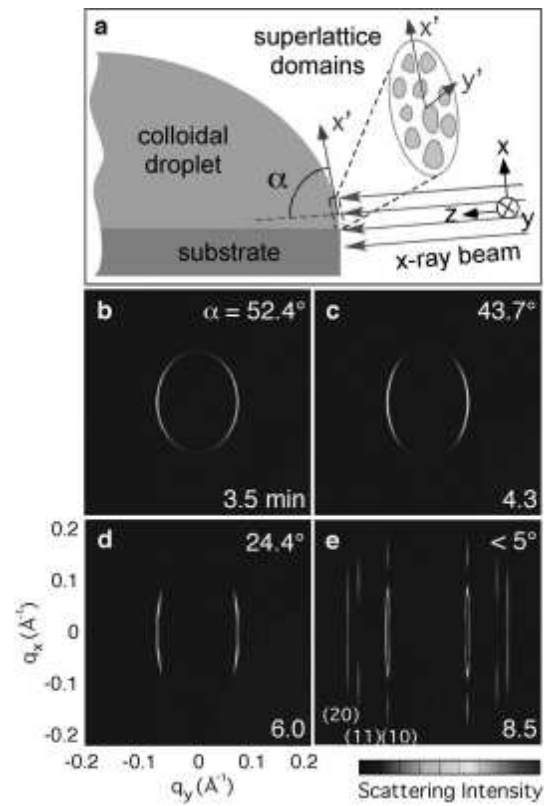}}
\caption{\label{fig3}Simulation results of the SAXS experiments.
(a) Schematic cross sectional diagram of the x-ray beam and the
colloidal droplet. (b)-(e) Simulation results based on a monolayer
of 2D domains oriented at an angle $\alpha$ with respect to the
incident x-ray beam at the liquid-air interface. The time in each
frame corresponds to the real time used in our experiments.}
\end{figure}

The formation of the 2D NCSs can be understood using a kinetic
"crushing" model proposed by Nguyen and Witten \cite{Ngu}. The
diffusion constant ($D$) of the nanocrystals with a hydrodynamic
radius R (5.1 nm) can be estimated using the Stokes-Einstein
relation $(D=k_{B}T/6\pi\eta R)$ and the viscosity of
dodecanethiol $(\eta=2.98cp$), assuming that toluene completely
evaporates near the liquid-air interface. Since the major change
in scattering patterns in Figure 2 occurs at a time scale of one
minute, we can compare the diffusion distance of nanocrystals in
one minute with the change of liquid-air interface during the same
time interval. Performing a random walk, the average distance a
nanocrystal can diffuse within one minute is
$\sqrt{<r^{2}>}=\sqrt{6Dt}\approx 70\mu m$ . Since the air-liquid
interface decreases faster than 70 $\mu$m/min during the initial
evaporation, as measured by monitoring the thickness change of the
droplet in real-time, nanocrystals will accumulate at the 2D
liquid-air interface and eventually induce 2D crystallization.

The exponential increase of the scattering intensity (Figure 2e)
also supports the interface "crushing" model, in contrast to the
power law behavior in the conventional 2D diffusion coarsening
mechanism \cite{Lo}. In the "crushing" case, the domain growth
mainly occur through incorporating nanocrystals impinging the
domain surface from below when it collapses with the evaporating
liquid-air interface. The collected nanocrystals either fill the
vacant sites or move to the edge of the domain. The growth rate of
a domain with size $S$ is $\rm{d}S/\rm{d}$t$\sim$$nSv$ , where $n$
is density of nanocrystal and $v$ is the collapsing rate of
interface. We note approximately $n$ and $v$ remains constant
during most of the evaporation process. The "crushing" effect thus
leads to an exponential growth of domain size, which is a much
more dominant process than the power-law growth by 2D diffusion at
the interface.

One prediction of this kinetic "crushing" model is that the
formation of the 2D NCSs should be affected by the initial
evaporation rate. A slower initial evaporation rate could allow
the nanocrystals to diffuse away from the interface before the
nanocrystal concentration reaches its critical value for 2D
crystallization. The formation of superlattices will then occur in
the bulk of the droplet once the overall concentration of
nanocrystals in the droplet exceeds a critical concentration for
3D nucleation. Figure 4 shows the time evolution of scattering
patterns when the droplet is evaporated at approximately 0.22mg
per minute, much slower than the ambient condition. Indeed, 3D
superlattices form and persist during the entire evaporation
process. Using the same colloidal solution, but through varying
the evaporation rate, both the 2D NCSs and 3D NCSs formation can
be reproduced consistently.

\begin{figure}
\scalebox{0.40}{\includegraphics{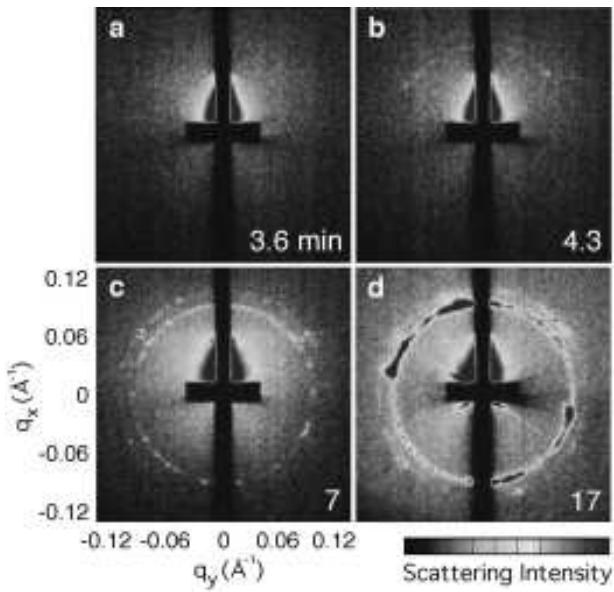}} \caption{\label{fig4}In
situ scattering patterns obtained during a slow evaporation of the
colloid droplet.  In this experiment, the average diameter of
nanocrystals is 5.8 nm determined by TEM. These patterns were
recorded by the CCD detector.}
\end{figure}

Our measurements suggest the kinetics of evaporation can strongly
affect the structure of NCS formation. 2D NCSs can form
exponentially at the liquid-air interface when the evaporation is
fast enough to induce nanocrystal accumulation at the liquid-air
interface, whereas under a much slower evaporation condition,
nanocrystals can diffuse away from the interface and 3D NCSs can
eventually form inside the droplet. The kinetics driven
self-assembly process observed in our nanocrystal colloidal
system, might also play an important role in other complex
systems. Exploring these systems using real-time x-ray scattering
techniques should enable us to better control the organization of
nanoscale building blocks, and also study many interesting
phenomena, such as particle and domain dynamics and
order-to-disorder phase transition.

We thank T. T. Nguyen, T. A. Witten, T. P. Bigioni, H. M. Jaeger
and C. M. Sorensen for extensive discussions. This work was
supported by the U.S. Department of Energy (DOE), BES-Materials
Sciences, under Contract \#W-31-109-ENG-38, by DOE Center for
Nanoscale Materials, and by the University of Chicago - Argonne
National Laboratory Consortium for Nanoscience Research (CNR). The
use of APS is supported by Office of Science of DOE.

\end{document}